\documentclass[preprint,amsmath,amssymb,longbibliography]{revtex4}
\usepackage{txfonts}
\usepackage{amsfonts}
\usepackage{amsmath}
\usepackage{graphicx} 
\usepackage{overpic}
\usepackage{color}
\usepackage{url}
\usepackage{enumerate}
\usepackage{algorithm}
\usepackage{algpseudocode}

\newcommand{\be}{\begin{eqnarray}}
\newcommand{\ee}{\end{eqnarray}}

\begin{document}

\title{Human Mobility in Large Cities as a Proxy for Crime}

\author{Carlos Caminha$^{1}\footnote{Correspondence to:
caminha@edu.unifor.br}$,Vasco Furtado$^{1}$,
Tarcisio H. C. Pequeno$^{1}$, Caio Ponte$^{1}$, Hygor
P. M. Melo$^{2,3}$, Erneson A. Oliveira$^{1,3}$,
Jos{\'e} S. Andrade Jr.$^{3}$}

\affiliation{$^1$ Programa de P{\'o}s Gradua{\c c}{\~a}o em Inform{\'a}tica
Aplicada, Universidade de Fortaleza, Fortaleza, Cear{\'a}, Brasil.\\ $^2$ Departamento de Ensino, Instituto Federal de Educa{\c
c}{\~a}o, Ci{\^e}ncia e Tecnologia do Cear{\'a}, Crate{\'u}s, Cear{\'a},
Brasil.\\ $^3$ Departamento de F{\'i}sica, Universidade Federal do
Cear{\'a}, Fortaleza, Cear{\'a}, Brasil.}

\date{\today}

\begin{abstract}
We investigate at the subscale of the neighborhoods of a highly populated city
the incidence of property crimes in terms of both the resident and the floating
population. Our results show that a relevant allometric relation could only be
observed between property crimes and floating population. More precisely, the
evidence of a superlinear behavior indicates that a disproportional number of
property crimes occurs in regions where an increased flow of people takes place
in the city. For comparison, we also found that the number of crimes of peace
disturbance only correlates well, and in a superlinear fashion too, with the
resident population. Our study raises the interesting possibility that the
superlinearity observed in previous studies [Bettencourt {\it et al.}, Proc.
Natl. Acad. Sci. USA {\bf 104}, 7301 (2007) and Melo {\it et al.}, Sci. Rep.
{\bf 4}, 6239 (2014)] for homicides versus population at the city scale could
have its origin in the fact that the floating population, and not the resident
one, should be taken as the relevant variable determining the intrinsic
microdynamical behavior of the system.
\end{abstract}

\maketitle

\section*{Introduction}

The dynamics of crime and the impact of social relations on the increase of
violence has been the object of study in several areas such as Social Sciences
\cite{michael1933}, Criminology \cite{cohen1979, felson2002, clarke1993},
Computing \cite{guedes2014, wang2013, kiani2015}, Economics \cite{wang2005} and
Physics \cite{melo2014, alves2013a, alves2013b}. When, in the 1950s, Naroll and
Bertalanffy \cite{von1963} utilized the concept of allometry – a term originally
coined in the field of Biology to describe scaling laws, {\it e.g.}, the
relationship between mass and metabolic rate of organisms \cite{kleiber1932} –
so as to adapt it to the social context, a particularly promising line of
research was opened, which today arouses interest of scientists from
wide-ranging areas \cite{oliveira2014, bettencourt2007, melo2014}. More
recently, Bettencourt {\it et al.} \cite{bettencourt2007} revealed that
allometric relationships are statistically present in many aspects of city
infrastructures and dynamics. In particular, they observed a characteristic
superlinear relation between the number of serious crimes and the resident
population in the United States (US) cities which clearly denotes the an
intricate social mechanism behind the dynamics of violence. More recently, Melo
{\it et al.} \cite{melo2014} provided strong quantitative evidence showing that
an entirely similar behavior can also be observed in Brazil.

Despite the importance of the results presented in the aforementioned studies,
their impacts on urban planning, more specifically, on the development of public
safety policies, are limited due to its purely descriptive nature, which
prevents a deeper understanding of the organic causes leading to such a
disproportional behavior. In this way, a microdynamical approach based on the
interactions between local neighborhoods within metropolitan areas certainly
represents a more realist view to the problem. Our research motivation is aimed
at elucidating issues related to the understanding of the impacts of the
influence of social relations on Crime. In Criminology, the role of urban space
and its social relations has been previously emphasized to explain the origin of
Crime \cite{cohen1979, clarke1993}. Particularly, the theory of routine
activities, proposed by Cohen and Felson \cite{cohen1979}, states that crimes,
more specifically property crimes, such as robbery and theft, occur by the
convergence of the routines of an offender, motivated to commit a crime, and an
unprotected victim. In this context, can we explain the occurrence of crimes in
different areas of the city based on the current population present in the
corresponding urban sub-clusters? Is this effective population equally important
for any type of crime? How can we systematically delimit the boundaries of these
local neighborhoods so that the social influence is accounted for in a
consistent way?

In order to answer these questions, we used actual georeferenced data of crimes
committed, and of resident and floating populations for census tracts in the
city of Fortaleza, Brazil. The concept of resident population has been widely
used to understand the effects that the growth of major cities has on social and
environmental indicators. In particular, Bettencourt {\it et al.}
\cite{bettencourt2007} showed that the number of homicides scales superlinearly
with the population of cities in the US. Subsequently, Melo {\it et al.}
\cite{melo2014} confirmed this behaviour for Brazilian cities, but also
demonstrated that suicides scale sublinearly with their resident populations.
Additionally, Oliveira {\it et al.} \cite{oliveira2014} found a superlinear
allometric relationship between the resident population and $CO_2$
emissions; this study also raised the issue that the allometric exponents may
undergo endogenous trends depending on the respective definition of urban
agglomerates. Regarding the floating population, it takes into account the
complex dynamics of urban mobility, {\it i.e.}, it has a transient
characteristic within the city. To measure social influence quantitatively in
urban sub-clusters, we delimited their boundaries beyond the mere administrative
divisions, {\it e.g.} division by neighborhoods or census tracts, by using the
City Clustering Algorithm \cite{makse1998, rozenfeld2008, giesen2010,
rozenfeld2011, duranton2013, gallos2012, duranton2015, eeckhout2004}. This model
defines these boundaries by population density and the level of commuting
between areas of the city.

In the present study, we identified that the incidence of property crimes has a
superlinear allometric relationship, with the floating population in certain
areas of the city. This result implies that the increased flow of people in a
particular area of the city will take place at the cost of a proportionally
greater rate of property crime happening in the region. More important, this
superlinear behavior at the subscale of the city neighborhood provides a
plausible explanation for the allometry of serious crimes found in
\cite{bettencourt2007,melo2014}. Precisely, the floating population being
systematically larger that the resident one should lead to the disproportional
behavior observed for serious crimes and (resident) population at the city
scale. We also found a superlinear allometric relationship between the number of
crimes of disturbing the peace and the resident population. This result shows
that the effect of social influence must be adequately correlated with resident
or floating population, depending on the type of crime considered.

\section*{Datasets}

\begin{figure}[!ht]
\includegraphics[width=1.0\textwidth]{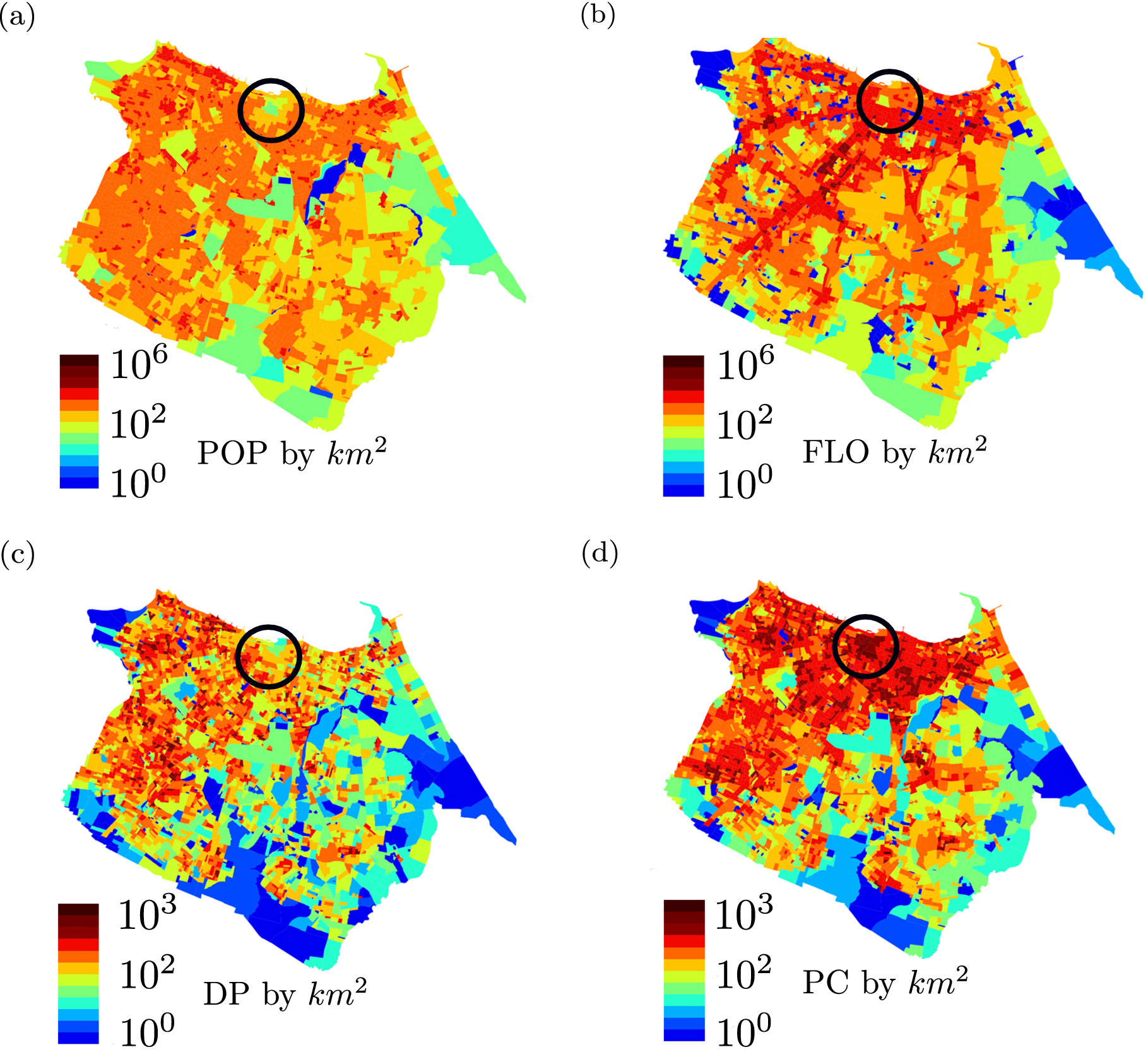}
\caption{{\bf The density maps for the city of Fortaleza (Brazil) in logarithmic
scale.} (a) The resident population (POP) by square kilometers
($\mathrm{km}^2$). (b) The floating population (FLO) by $\mathrm{km}^2$. (c) The
disturbing the peace (DP) complaints by $\mathrm{km}^2$. (d) The property crimes
(PC) by $\mathrm{km}^2$. The black circle highlights the downtown area of the
city. This region has a low density of residents and disturbing the peace calls,
and is dense in the flow of people and property crimes.}
\label{fig1}
\end{figure}

In order to quantify the effects of the social influence on the police calls
within a large metropolis, we used three georeferenced datasets for the
Brazilian city of Fortaleza: From the first, we obtain the {\it resident
population} (POP), defined by the number of residents per {\it census tract} –
Administrative territorial unit established for the purposes of cadastral
control – and provided by the Brazilian Institute of Geography and Statistics
(IBGE) \cite{ibge2016}. In all, Fortaleza has 3018 census tracts, with
approximately 2,400,000 residents, spread over an area of 314 square kilometers
($\mathrm{km}^2$) in year 2010. From the second, we estimate the {\it floating
population} (FLO) for each census tract through a flow network built on data of
the bus system provided by the Fortaleza's city hall for the year 2015
\cite{dadosAbertosFortaleza2016}. FLO was measured by the number of people who
pass through a census tract in one day. The city of Fortaleza has 2034 buses
circulating along 359 bus lines serving approximately 700,000 people who use the
city’s mass transit system on a daily basis. In the case of Fortaleza, buses
still represent the main means of public transportation. The process of
generating of the flow network will be detailed in the Supporting Information
(see the S1 Appendix). Finally, we obtain the {\it Crime dataset} from
the Integrated Coordination Office of Public Safety Operations (CIOPS)
\cite{ciops2016}, provides the geographic locations of 81,911 calls to the
``190'' (phone number for emergency) service about property crimes (PC) and
53,849 calls to the same service about disturbing the peace (DP). These calls
were made to police between August 2005 and July 2007. The color maps show in
Figs \ref{fig1}(a)-\ref{fig1}(d) provide the local density in logarithmic scale
of POP, FLO, DP and PC, respectively, for the city of Fortaleza. We can clearly
identify stronger correlations between POP and DP as well as between FLO and PC.
In special, there is an evident higher incidence of hot spots in (b) and (d)
than in (a) and (c). Also, at the downtown area, highlighted by black circles on
each map, the high density of FLO is compatible with the high rates of PC rate,
while low POP densities seem to explain the low frequency of DP complaints.

\begin{figure}[!ht]
\includegraphics[width=1.0\textwidth]{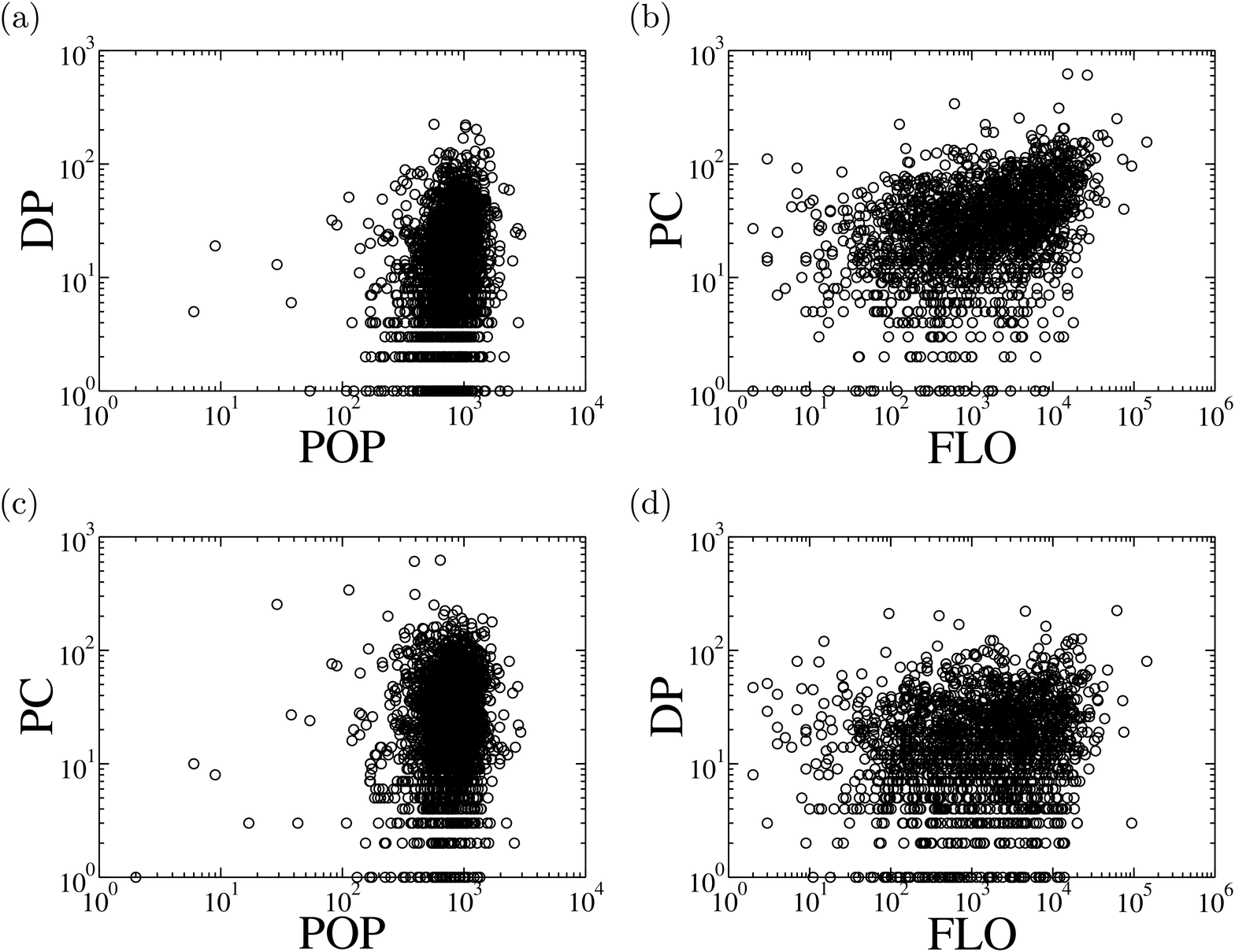}
\caption{{\bf Scattering plots by census tracts.} (a)-(d) All the plots between
resident population (POP) and floating population (FLO) with police calls for
disturbing the peace (DP) and for property crimes (PC) show uncorrelated
behavior with the determination coefficients \cite{rawlings2001, montgomery2015}
$R^2 < 0.15$.}
\label{fig2}
\end{figure}

In spite of seeming trivial to suggest that there are correlations between POP
and DP, as well as FLO and PC from the density maps shown in Fig \ref{fig1},
the respective scattering plots show uncorrelated behaviors (Fig \ref{fig2}).
Actually, we conjecture that such correlations exist indeed. The most census
tracts have a small area, sometimes the size of one city block, and it is likely
that such an agglomeration scale is insufficient to capture the correlations and
therefore reveal the impact of social influence on DP and PC. Based on this
hypothesis, we considered coarse-grainning these spatial properties of Fortaleza
into clusters, using the census tracts grid as a maximal resolution base. This
clustering aims to find the boundaries of the flow of people and the residential
areas of the city.

\section*{Methods}

\begin{figure}[!ht]
\includegraphics[width=1.0\textwidth]{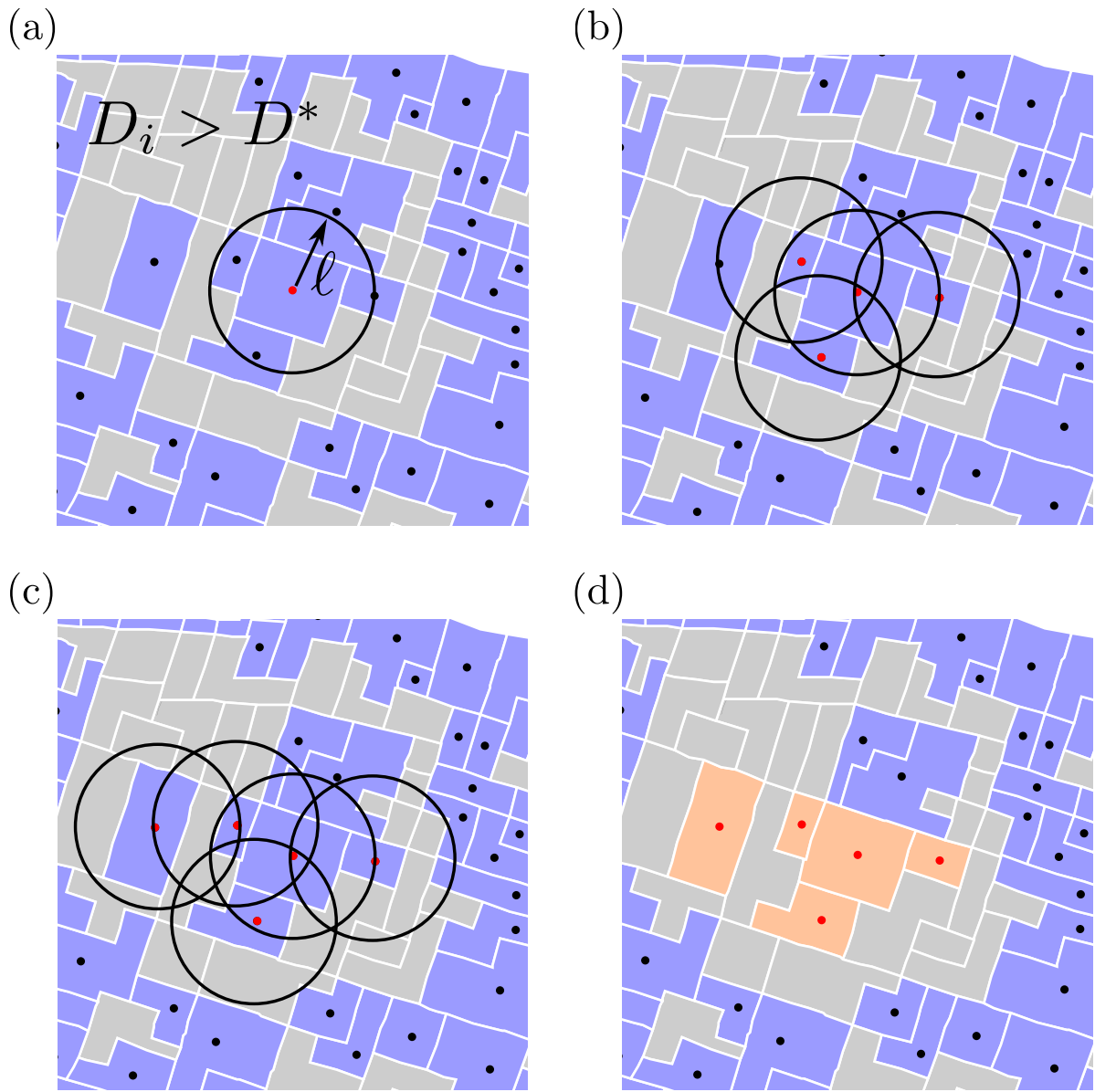}
\caption{{\bf The scheme of the City Clustering Algorithm (CCA).} Each polygon
represents a clustering unit, specifically in our case, they represent census
tracts. The light blue polygons are candidates for clustering $(D_i > D^*)$; in
contrast, the gray polygons cannot be clustered ($D_i \le D^*$). (a) The red dot
represents the geometric center of the $i\mathrm{-th}$ census tract and the
black circle with radius $\ell$ seeks neighbors belonging to the same cluster.
(b)-(c) The same search operation is made for the other census tracts and is
done until there are no more neighbors within the radius of operation. (d) The
algorithm finishes running and the cluster is found.}
\label{fig3}
\end{figure}

To define the boundaries beyond administrative delineations, we considered the
notion of spatial continuity through the aggregation of census tracts that are
near one another using the City Clustering Algorithm (CCA) \cite{makse1998,
rozenfeld2008, giesen2010, rozenfeld2011, duranton2013, gallos2012,
duranton2015, eeckhout2004}. The CCA constructs the population boundaries of an
urban area considering two parameters, namely, a population density threshold,
$D^*$, and a distance threshold, $\ell$. For the $i\mathrm{-th}$ census tract,
the population density $D_i$ is located in its geometric center; if $D_i > D^*$,
then the $i\mathrm{-th}$ census tract is considered populated. The length $\ell$
represents a cutoff distance between census tracts to consider them as spatially
contiguous, {\it i.e.}, all of the nearest neighboring census tracts that are at
distances smaller than $\ell$ are clustered. Hence, a cluster made by the CCA is
defined by populated areas within a distance less than $\ell$, as seen
schematically in Fig \ref{fig3}. Previous studies
\cite{oliveira2014,duranton2013, duranton2015} have demonstrated that the
results produced by the CCA can be weakly dependent on $D^*$ and $\ell$ for some
range of parameter values. Here $\ell$ will be quantified in meters (m) and
$D^*$ in inhabitants by $\mathrm{km}^2$.

Although the algorithm begins collating an arbitrary seed census tract, it does
not produce distinct clusters when varying this seed; the two factors that are
responsible for clustering behavior are the parameters $\ell$ and $D^*$.

\begin{figure}[!ht]
\includegraphics[width=1.0\textwidth]{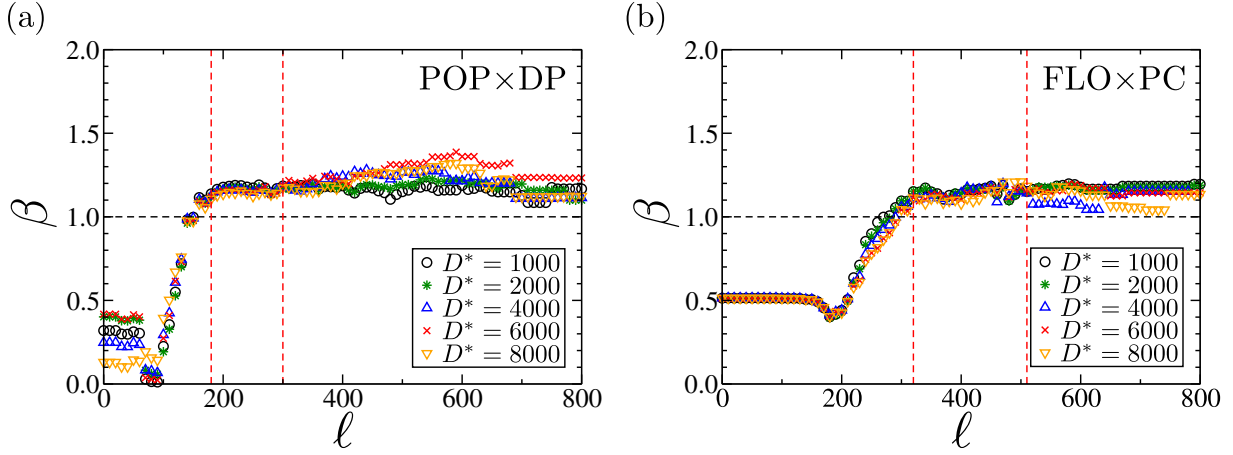}
\caption{{\bf Behavior of exponent $\boldsymbol\beta$ by varying the parameters
of the City Clustering Algorithm (CCA), $\boldsymbol\ell$ and $\boldsymbol
D^*$.} (a) The variation of $\beta$ in correlation between the resident
population (POP) and the disturbing the peace (DP) complaints is illustrated;
(b) The variation is illustrated for correlations between the floating
population (FLO) and the property crimes (PC). Both in (a) and (b), the x-axis
represents $\ell$, and this parameter was varied from 0 to 800 meters (m)
(moment when the largest cluster consumes nearly the entire city); exponent
$\beta$ is shown on the y-axis. The colors of the lines represent the variation
of the parameter $D^*$, which corresponds to the resident population density in
(a) and the floating population density in (b); this parameter was varied from
1000 to 8000. It was not necessary to use values larger than 8000 because many
census tracts start being discarded and the CCA can no longer form clusters. The
graphs also show red dashed lines; between these lines is highlighted the range
where, regardless of the parameterization, the exponent $\beta$ has smaller
ranges of variation. Finally, the dotted black line highlights exponent $\beta =
1$, in which the relationship between variables is isometric, in both graphs the
exponent oscillates to low values of $\ell$; in (a), the relationship is
superlinear starting at $\ell \ge 180$ m; however in (b), superlinearity appears
at $\ell \ge 320$ m.}
\label{fig4}
\end{figure}

In order to determine the effect of the parameterization on the value of the
exponent $\beta$, we sought a range within the parameters where $\beta$ has low
sensitivity to this variation. Fig \ref{fig4} shows the behavior of exponent
$\beta$ in function of the variation of the CCA parameters. As shown in Figs
\ref{fig4}a and \ref{fig4}b, the value of the parameter $\beta$ obtained from
the least-square regressions to the data of POP against DP remains practically
insensitive to the CCA parameters in the range $180 \ge \ell \ge 300$,
regardless of the values of $D^*$ adopted in the estimation process. Moreover,
the resulting average $\beta=1.17\pm0.06$ provides strong evidence to support a
superlinear type of relation between these two variables. An entirely similar
behavior can be observed for FLO against DP, but now the exponent $\beta$
remains practically invariant within the range $320 \geq \ell \geq 510$. The
resulting average value of $\beta=1.14\pm0.04$ also indicates the presence of a
superlinear allometric relation.

At this point, it is necessary to determine a suitable criteria for selecting an
adequate value of the parameter $D^*$. While lower values of $\ell$ lead to the
formation of a larger number of CCA clusters, reduced values of $D^*$ tend to
eliminate fewer census tracts from the map, thereby including a larger portion
of in the population under analysis. Here we propose that a proper choice of
$D^*$ would be associated with a more homogeneous spatial distribution of the
population \cite{oliveira2014}. More precisely, we seek for CCA clusters whose
areas should scale as close to isometrically as possible with the population
data, namely, as $Y=aX^{\alpha}$, with $\alpha \approx 1$, where $X$ is the
population, $Y$ is the area (ARE) of the clusters $\mathrm{km}^2$, and $a$ is
constant.

\begin{figure}[!ht]
\includegraphics[width=1.0\textwidth]{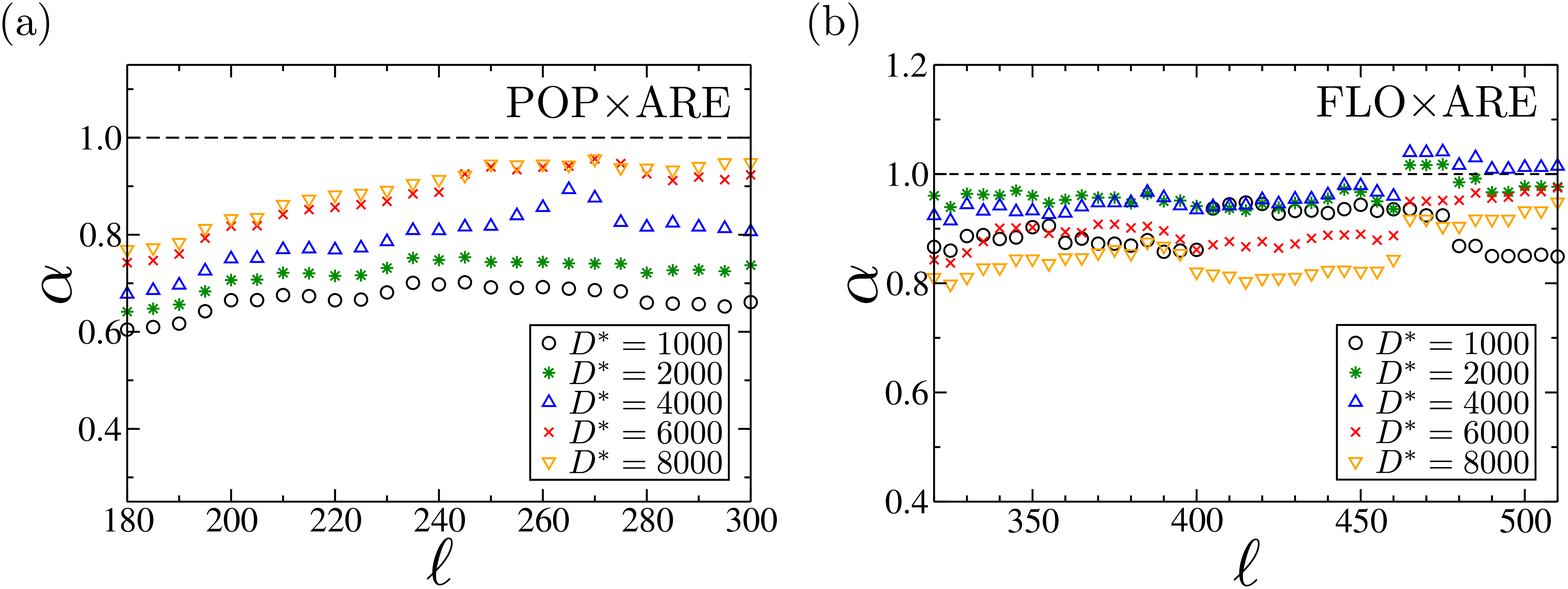}
\caption{{\bf Behavior of exponent $\boldsymbol\alpha$ when varying the City
Clustering Algorithm (CCA) parameters $\boldsymbol\ell$ and $\boldsymbol D^*$.}
(a) The variation of $\alpha$ in correlations between the resident population
(POP) and the area (ARE) in square kilometers ($\mathrm{km}^2$) of clusters
discovered with the CCA. (b) The variation for correlations between the floating
population (FLO) and ARE. In (a) and (b), the x-axis represents the parameter
$\ell$, and the y-axis represents the exponent $\alpha$. The line colors
represent the variation of the parameter $D^*$.}
\label{fig5}
\end{figure}

In order to follow the procedure previously described, we obtain from
Fig \ref{fig5}a that $\ell=270$ and $D^{*}=6000$ correspond to the pair of CCA
parameter values leading to the closest to isometric relation found between ARE
and POP. In the case of ARE and FLO, the values are $\ell=320$ and
$D^{*}=2000$, as depicted in Fig \ref{fig5}b (see the S2 Appendix).

\section*{Results}

\begin{figure}[!ht]
\includegraphics[width=1.0\textwidth]{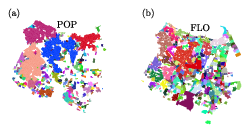}
\caption{{\bf The City Clustering Algorithm (CCA) applied to resident population
(POP) and to floating population (FLO).} Each color represents a cluster; the
white areas correspond to census tracts that were not grouped because they
have $D_i < D^*$. (a) The population density was used in order to find the
boundaries of the clusters with $\ell$ = 270 m and $D^*$ = 6000 resident people
per $\mathrm{km}^2$. (b) The division found by considering urban mobility is
shown; the map illustrated here was generated for $\ell$ = 320 m and $D^*$ =
2000 floating people per km$^2$ in one day in Fortaleza.}
\label{fig6}
\end{figure}

The census tracts were grouped, using the CCA, by POP and FLO (Fig \ref{fig6}).
In the Fig \ref{fig6}a, the division achieved by POP is illustrated; the city
was divided using $\ell$ = 270 m and $D^*$ = 6000 resident people per
$\mathrm{km}^2$. In the Fig \ref{fig6}b, we show the division achieved by FLO,
using $\ell$ = 320 m and $D^*$ = 2000 floating people per $\mathrm{km}^2$ per
day. We emphasize that there are bigger gaps in the POP map (Fig \ref{fig6}a)
than in the FLO map (Fig \ref{fig6}b). The reason for such behavior is the fact
that Fortaleza has commercial regions, {\it i.e.}, regions where there is not a
large presence of residents. Regarding floating people, there are people moving
practically throughout the entire city, both in commercial areas and in
residential areas.

\begin{figure}[!ht]
\includegraphics[width=1.0\textwidth]{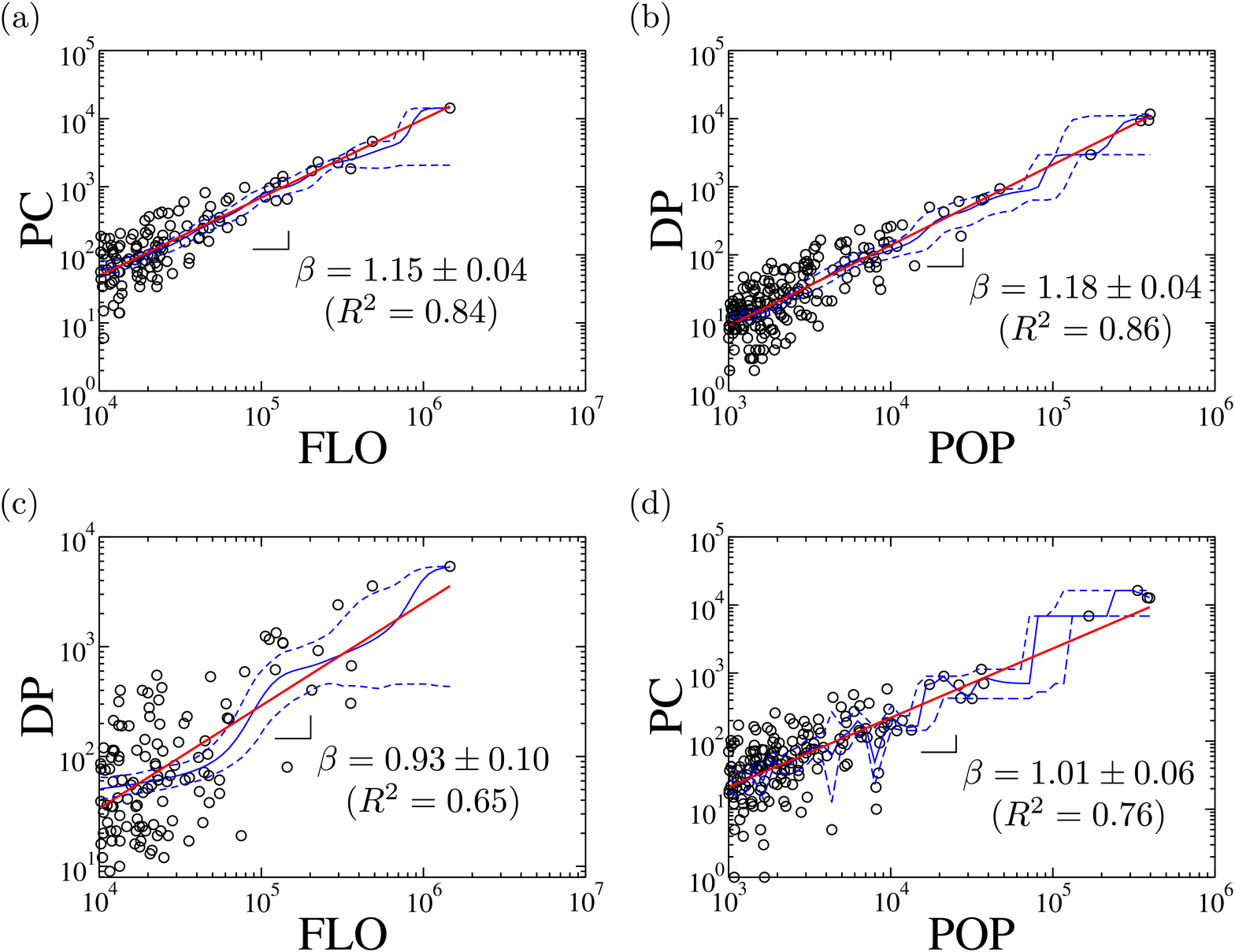}
\caption{{\bf Scattering plots by City Clustering Algorithm (CCA).} The red
lines represent the simple linear regressions applied to the data, the blue
continuous lines represent the Nadaraya-Watson method
\cite{nadaraya1964,watson1964} and the blue dashed lines delimit the 95\%
confidence interval (CI) estimated by bootstrap. (a) A superlinear relationship
was found, with exponent $\beta = 1.15 \pm 0.04$, between the floating
population (FLO) and the property crimes (PC). (b) A superlinear relationship
was also found between the resident population (POP) and the disturbing the
peace (DP) complaints, with exponent $\beta = 1.18 \pm 0.04$. (c)-(d) The
scatering plots of DP with FLO and PC with POP show an isometric relation was
found between the variables, but with lower correlations than (a) and (b). The
$R^2$ is defined as determination coefficient \cite{rawlings2001,
montgomery2015}.}
\label{fig7}
\end{figure}

As compared to the results shown in Fig \ref{fig2}, the application of the CCA
to the data discloses a rather different scenario for the correlations among the
variables investigated here. First, as shown in Fig \ref{fig7}a and
\ref{fig7}b, superlinear relations in terms of power laws, $Y = aX^\beta$, are
revealed between PC and FLO as well as between DP and POP, with exponents $\beta
= 1.15 \pm 0.04$ and $\beta = 1.18 \pm 0.04$, respectively. In contrast, the
relations obtained between DP and FLO and PC and POP are closer to isometric
(linear), with exponents $\beta = 0.93 \pm 0.10$ and $\beta = 1.01 \pm 0.06$
respectively, although the low values of the corresponding determination
coefficients indicate that these results should be interpreted with caution.

\section*{Conclusion}

In this paper, we define a methodology to understand the impact that society has
on calls to the police finding the boundaries of social influence in a large
metropolis through the analysis of georeferenced datasets for the city of
Fortaleza, Brazil. We used the CCA, a clustering algorithm, in order to define
intracity clusters, {\it i.e.} spatially contiguous populated areas within a
cutoff distance. The volume of social influence was measured in various
localities of the city based on the presence of residents and the flow of
passers-by through the census tracts. Unlike intercity studies, where social
influence was measured only by the presence of residents, we propose that,
within a city, urban mobility should be considered to understand the dynamics of
social indicators, such as criminal activity. Our results show that the
incidence of property crimes grows superlinearly as a power law with the
floating population, with allometric exponent $\beta = 1.15 \pm 0.04$.
Therefore, the increase of the flow of people in a region of the city leads to a
disproportionally higher number of property this type of crime. Our results are
in agreement with the routine activities hypothesis \cite{cohen1979}, which
states that a crime occurs by the convergence of the routines of three agents,
specifically: The presence of a motivated offender, the presence of an
unprotected victim, and the absence of a guardian able to prevent the
transgression. In other words, the regions with higher incidence of floating
population potentiate the meeting of the routines of these three agents. This
result is in clear contrast with the incidence of crimes related with peace
disturbance, where an allometric relation can also be detected, but with the
resident population (POP) instead. We hope the our results could shed some light
on the understanding of Crimes inside urban areas, as well as assist eventual
Violence mitigation policies.

The findings described in this paper bring alternatives to implementing
innovative practices to decision makers within cities. The most obvious of these
relates to the fact that, by showing the correlation of different types of
crimes with the home population but also with the floating population, it is
also clear that the police force allocation strategies should be implemented via
the analysis of different cluster configurations that depend on the type of
crime. For example, the allocation of community policing, more appropriate to
resolve conflicts that potentially can emerge from the disturbance of people's
peace, must be planned from a cluster configuration and a hot spot analysis that
were produced from the perspective of the density of resident population. When
it is necessary to establish a policy for allocating a uniformed police in order
to mitigate crimes against property, the allocation of the police force must be
conducted from an analysis from the movement of people.

In addition to these police allocation strategies, the results described herein
provide important indicators for the creation of public policies for land use
and environmental design in general. Work in this line has been developed such
as in Ref. \cite{taylor1996} where a framework has been proposed to associate
the physical spaces and the feeling of safety as well as \cite{brantingham1981},
who launched the environmental criminology putting focus of criminological study
on environmental or context factors that can influence criminal activity. These
include space (geography), time, law, offender, and target or victim. These five
components are a necessary and sufficient condition, for without one, the other
four, even together, will not constitute a criminal incident. The discovery
demonstrated in the article that there is a superlinear relationship between
crime and population (resident or floating) in clusters within cities
strengthens the claim that changes in urban form can lead to reduced crime as
discussed in Ref. \cite{kinney2008}.

\section*{Supporting Information}

\section*{S1 Appendix. The urban mobility data processing}

The urban mobility system of a large city is composed of several interconnected
networks, such as subway, bus, bicycle, taxicab, and private vehicle networks.
Buses are the main means of transportation for the most inhabitants in the city
of Fortaleza, being used by about 700,000 people daily. Taking this fact into
account, we assumed that the urban mobility within the city can be represented
by the use of the bus system. Thus, the trajectories of bus users will be used
to infer the floating population at the different points of the city.

In order to understand the people flow throughout the city, we used four
spatio-temporal datasets related to Fortaleza's bus network, which are: bus
stops; bus lines; GPS tracking of vehicles; and ticket card validation. All
datasets refer to a normal business day, in fact, the March 11th, 2015 - a
Wednesday. In total, Fortaleza has 4,783 bus stops served by 2,034 buses along
359 different routes. The integrated transportation model adopted by Fortaleza
City Hall, called {\it Bilhete {\'U}nico}, allows the registred users to make a
bus transfer anywhere in the city, as long as it is within two hours since the
last validation of their ticket card. The validation process is understood as
the act of the user swiping his/her ticket card at the turnstile on the bus or
at the bus terminal. Usually, such procedure happens at the beginning of the
trip, since the turnstile is close to the bus entrance in Fortaleza. In this
context, we are able to define the Origin-Destination Matrix (ODM) for
Fortaleza's bus network through the following hypoteses \cite{caminha2016a}: We
can assume that an user's origin point could be represented by the earliest of
all first daily ticket validations in the interval of two weeks before March
11th, as well as an user's destination point could also be represented by the
earliest of all last daily ticket validations in the interval of one month
before March 11th, bearing in mind that an user could have different destination
points along the week, {\it i.e.} we have to analyze Mondays with Mondays,
Tuesdays with Tuesdays, and so on. Hence, we estimated the origin-destination
pair for about 40\% of the bus users representing the overall behavior of the
urban mobility in Fortaleza.

Finally, we supposed that the trajectories of bus users are defined by the
composition of routes of the buses took by them between their origin-destination
pair \cite{caminha2016b}. In this context, we describe the trajectories of bus
users as a directed graph $G(V,E)$, where $V$ and $E$ are the set of vertices
$v$ and edges $e$, respectively. An edge $e$ between the vertices $v_i$ and
$v_j$ is defined by the ordered pair $(v_i,v_j)$. In our approach, the vertices
represent bus stops and the edges represent the demand of bus users between two
consecutive bus stops. For each vertice $v_i$, we defined a weighted function
$d(v_i)$ as the sum of the users passing by $v_i$. Thus, we calculated the
floating population as the sum of all $d(v_i)$ within each census tracts.

\section*{S2 Appendix. The isometric relation between population and area}

\begin{figure}[!h] \includegraphics[width=1.0\textwidth]{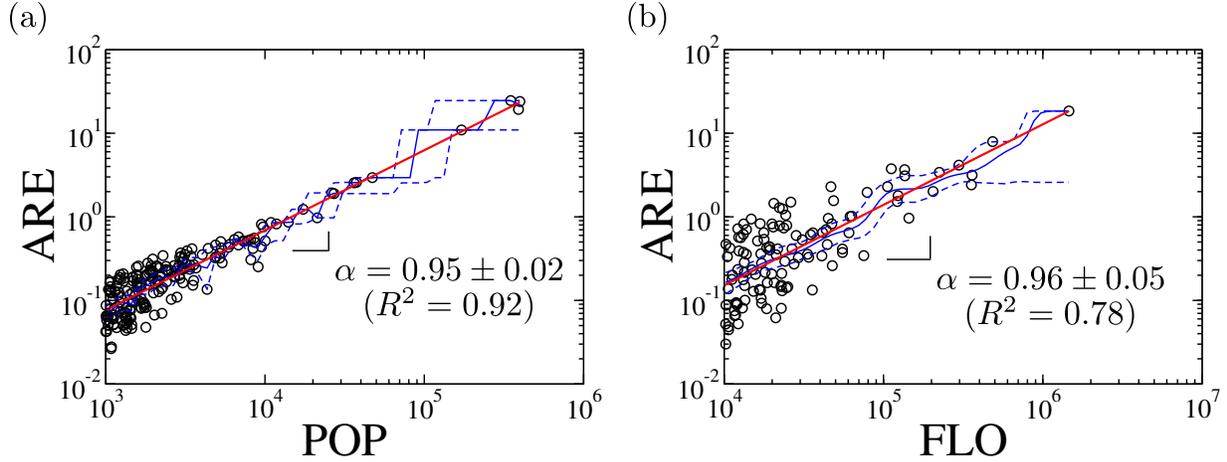}
\caption{{\bf The isometric relation between resident (POP) and floating
population (FLO) with area in square kilometers ($\mathrm{km}^2$).} (a) The
correlation between POP and the area of the clusters (ARE) calculated using City
Clustering Algorithm (CCA) where $D^*=6000$ residents per $\mathrm{km}^2$ and
$\ell=250$ meters (m). (b) The correlation between FLO and ARE calculated using
the CCA for $D^*=2000$ floating people per $\mathrm{km}^2$ and $\ell = 320$ m.
The red line shows the Ordinary Least Square (OLS) regression applied to the
logarithm of the data \cite{rawlings2001,montgomery2015}, and the blue
continuous line indicates the Nadaraya-Watson kernel regression
\cite{nadaraya1964,watson1964}. Finally, the blue dashed lines delimit the 95\%
confidence interval estimated by 500 random bootstrapping samples with
replacement \cite{racine2004,li2004}. The $R^2$ is defined as the determination
coefficient \cite{rawlings2001, montgomery2015}.}
\label{S2_Appendix_Fig2}
\end{figure}

In the main text, we observed the emergence of a trend toward an isometric
relation between the populations (POP and FLO) of the CCA clusters and their
respective areas (ARE). The Fig \ref{S2_Appendix_Fig2} shows the closest
relations to an isometric behavior for both cases. The CCA parameters used to
define the clusters were $D^*=6000$ and $\ell=270$ for the POP case, illustrated
in Fig \ref{S2_Appendix_Fig2}a, and $D^* = 2000$ and $\ell = 320$ for the FLO
case (Fig \ref{S2_Appendix_Fig2}b).

\end{document}